\documentclass[amsmath,amssymb,amsfonts,12pt]{revtex4}
\usepackage{epsfig}

\def\ra{\rangle}
\def\la{\langle}

\begin{document}
\baselineskip22pt

\title{Relative entropy of entanglement of rotationally
invariant states}
\author{Zhen Wang}
\author{Zhixi Wang\footnote{Corresponding author: wangzhx@mail.cnu.edu.cn}}

\address{School of Mathematical Sciences, Capital Normal University, Beijing
100048, China}

\begin{abstract} We calculate the relative entropy of entanglement
for rotationally invariant states of spin-$\frac{1}{2}$ and
arbitrary spin-$j$ particles or of spin-1 particle and spin-$j$
particle with integer $j$. A lower bound of relative entropy of
entanglement and an upper bound of distillable entanglement are
presented for rotationally invariant states of spin-1 particle and
spin-$j$ particle with half-integer $j$.

\vskip 5mm

\noindent{\sl PACS:}  03.65.Ud, 03.67.-a

\noindent{\sl Keywords:} relative entropy, rotationally invariant
state

\end{abstract}

\maketitle

\section{Introduction}
Quantum entanglement has played a significant role in the field of
quantum information and quantum computation \cite{nilsen}. This
attracts an increasing interest in the study of quantification of
entanglement for any quantum state. Although a lot of entanglement
measures have been proposed for a generic mixed state, only partial
solutions such as entanglement of formation (EoF) for two qubits
\cite{hill,woot}, are known to give the closed forms for generic
bipartite states. In particular, symmetric states have elegant forms
to quantify their entanglement. For example, E.M. Rains \cite{rain1}
and Y.X. Chen {\it et. al} \cite{chen} calculate the relative
entropy of entanglement (REE) for maximally correlated state by use
of different ways. In addition, P. Rungta {\it et. al.} obtained
concurrence for isotropic states in \cite{caves} and K. Chen {\it
et. al} presented tangle and concurrence for Werner states in
\cite{ck}.

As invariant states under $SO(3)$ group have a relatively simple
structure and therefore have been investigated extensively in the
literature \cite{john1,john2, breuer1,breuer2,breuer3,chru3,augu}.
This family of symmetric states under $SO(3)$ group is called
rotationally invariant (RI) states. K.K. Manne and C.M. Caves
\cite{manne} derived an analytic expression for the EoF,
I-concurrence, I-tangle and convex-roof-extended negativity of RI
states of a spin-$j$ particle and spin-$\frac{1}{2}$ particle by
using K.G.H. Vollbrecht and R.F. Werner's method \cite{werner}. It
is known that the REE is one of the fundamental entanglement
measures, as relative entropy is one of the most important functions
in quantum information theory. Therefore, in this paper we apply
K.G.H. Vollbrecht and R.F. Werner's technique to derive the relative
entropy of entanglement for RI states of a spin-$\frac{1}{2}$
particle and arbitrary spin-$j$ particle or of a spin-1 particle and
a spin-$j$ particle with integer $j$.

This paper is organized as follows. In section \ref{pre} at first we
review the definition and some properties of REE and the
representations of RI states for two particles. Then we show the
simplified expression of REE on RI states. In section \ref{2xn} we
calculate the REE explicitly for RI state of spin-$\frac{1}{2}$ and
spin-$j$ particles. The REE is compared for different spin-$j$
particle. In section \ref{3xn} at first we obtain the relative
entropy of entanglement for RI state in the system of two spin-1
particels. Subsequently, we introduce a way to obtain the separable
state which minimizes the REE for RI state of the case $j_1=1,\
j_2=2$. Furthermore, this result can be extended to the case of
spin-1 particle and spin-$j$ particle with integer $j$. Thus we
obtain the REE for this family of RI states. Finally, a lower bound
of REE and an upper bound of distillable entanglement are presented
for RI states of spin-1 particle and spin-$j$ particle with
half-integer $j$. In section \ref{conclu} a few conclusions are
drawn.

\section{Preliminaries}\label{pre}
Throughout this paper we refer to $\mathcal S,\ \mathcal D$ and
$\mathcal P$ as the set of all states, separable states and positive
partial transposition (PPT) states, respectively. Relative entropy
\cite{vedral1,vedral2} of entanglement is defined as
\begin{eqnarray}\label{def}
 E_r(\rho)=\min_{\sigma\in{\mathcal D}}S(\rho\|\sigma)
=\min_{\sigma\in{\mathcal D}}{\rm tr}(\rho\ln\rho-\rho\ln\sigma).
\end{eqnarray}
There are some properties of REE as follows \cite{rain1,vedral2}:
\begin{enumerate}
\leftskip10pt\item[{(1)}] $E_r(\rho)\geq0$ with the equality
saturated iff $\rho$ is a separable state.
\item[{(2)}] Local unitary operations leave $E_r(\rho)$ invariant.
\item[{(3)}] $E_r(\rho)$ cannot increase under LOCC.
\item[{(4)}] For a pure state $\rho$ we have $E_r(\rho)=S(\rho_A)=
-{\rm tr}(\rho_A\ln\rho_A)$, where $S(\rho_A)$ is the entropy of
entanglement of $\rho_A={\rm tr}_B(\rho),\ {\rm tr}_B$ is a map of
operators known as the partial trace over system $B$.
\item[{(5)}] If $\sigma^*$ minimizes $S(\rho\|\sigma^*)$ over
$\sigma\in{\mathcal D}$ then $\sigma^*$ is also a minimum for any
state of the form $\rho_x=(1-x)\rho+x\sigma^*.$
\item[{(6)}] $E_r(x_1\rho_1+x_2\rho_2)\leq x_1E_r(\rho_1)+x_2E_r(\rho_2),
\ \hbox{where } x_1+x_2=1$ and $x_1,\ x_2$ are non-negative and
real.
\item[{(7)}] $E_r(\rho)\leq E_F(\rho)$, where $E_F(\rho)$ is EoF.
\item[{(8)}] $E_D(\rho)\leq E_\Gamma(\rho)\leq E_r(\rho),$ where
$E_D(\rho)$ is distillable entanglement and
$E_\Gamma(\rho)=\min_{\sigma\in{\mathcal P}}S(\rho\|\sigma),$
$\Gamma$ denotes the partial transpose operator.
\end{enumerate}

Now we recall the representations of RI states for two particles
with spins $j_1$ and $j_2$ and corresponding angular momentum
operators ${\hat{j}}^1$ and ${\hat{j}}^2$ \cite{john1,john2}.
Throughout this paper we will assume that $j_2\geq j_1$. The tensor
product ${\Bbb C}^{N_1}\otimes{\Bbb C}^{N_2}$ is the Hilbert space
of a system which is composed of spin-$j_1$ particle and spin-$j_2$
particle. The Hilbert space ${\Bbb C}^{N_1}$ of the first space is
spanned by the common eigenvectors $|j_1,m_1\ra$ of the square of
${\hat{j}}^1$
 and of ${\hat{j}}^1_z$, where
$N_1=2j_1+1$ and $m_1=-j_1,\ \cdots,\ +j_1.$ The Hilbert space
${\Bbb C}^{N_2}$ of the second space is spanned by the eigenvectores
$|j_2,m_2\ra$, where $N_2=2j_2+1$ and $m_2=-j_2,\ \cdots,\ +j_2$,
correspondingly. H.P. Breuer \cite{breuer1,breuer2} considered the
representation of RI states which employs the projection operators
$P_J=\displaystyle\sum_{M=-J}^{J}|JM\ra\la JM|$. Notice that here
$|JM\ra$ is the common eigenvector of the square of the total
angular momentum operator and of its $z$-component. RI states using
the representation can be written as
\begin{eqnarray}\label{state}
\rho=\frac{1}{\sqrt{N_1N_2}}\sum_{J=j_2-j_1}^{j_1+j_2}\frac{\alpha_J}
{\sqrt{2J+1}}P_J,
\end{eqnarray}
where the $\alpha_J$ are real parameters and $\sqrt{N_1N_2}$ and
$\sqrt{2J+1}$ are introduced as convenient normalization factors. In
order for $\rho$ to represent a density matrix the $\alpha_J$ must
be positive and normalized appropriately:
\begin{eqnarray}\label{con}
 \alpha_J\geq0,\qquad {\rm tr} \rho=\sum_J\sqrt{\frac{2J+1}{N_1N_2}}\alpha_J=1.
\end{eqnarray}
We denote the set of all vectors $\vec{\alpha}$ whose components
$\alpha_J$ satisfy the relations (\ref{con}) by $S^\alpha$. It is
obvious that $S^\alpha$ is isomorphic to the set of RI states and of
course a convex set. It is remarkable that the representation
(\ref{state}) is the spectral decomposition of $\rho$.

In the following we present the simplified expression of REE on RI
states. What makes the calculation of the REE easy for RI states is
the existence of a "twirl" operation \cite{werner2}, a projection
operator ${\bf P}$ that maps an arbitrary state $\rho$ to a RI state
${\bf P}(\rho)$ and that preserves separability, i.e., that maps
every separable state to a RI separable state. Since for a RI state
$\rho$ we have
\begin{eqnarray*}
S(\rho\|\sigma)\geq S({\bf P}(\rho)\|{\bf P}(\sigma)),
\end{eqnarray*}
 this guarantees that the minimum REE for a RI state is attained on
 another RI separable state \cite{werner}.
Hence, suppose
\begin{eqnarray*}
 \rho^*=\frac{1}{\sqrt{N_1N_2}}\sum_{J=j_2-j_1}^{j_1+j_2}
\frac{{\alpha_J}^*}{\sqrt{2J+1}}P_J
\end{eqnarray*}
be a RI separable state, one can show
\begin{eqnarray}\label{red}
E_r(\rho)=\min\sum_{J=j_2-j_1}^{j_1+j_2}\sqrt{\frac{2J+1}{N_1N_2}}\alpha_J({\rm
ln}\alpha_J-\ln\alpha_J^*),
\end{eqnarray}
where we utilize the fact that equation (\ref{state}) is the
spectral decomposition of RI state $\rho$.

\section{REE for $2\otimes N$ system}\label{2xn}
For a bipartite system consisting of a spin-$\frac{1}{2}$ particle
and a spin-$j$ particle, the RI state $\rho$ can be written as a
function of a single parameter $p$:
\begin{equation}\label{2n}
 \rho=\frac{p}{2j}\sum_{m=-j+\frac{1}{2}}^{j-\frac{1}{2}}
|j-\frac{1}{2},m\ra\la j-\frac{1}{2},m|
+\frac{1-p}{2j+2}\sum_{m=-j-\frac{1}{2}}^{j+\frac{1}{2}}
|j+\frac{1}{2},m\ra\la j+\frac{1}{2},m|.
\end{equation}
This equation is the spectral decomposition of $\rho$ with the
eigenvalues $p$ and $1-p$. From \cite{john1} we know that $\rho$ is
separable iff $p\leq\frac{2j}{2j+1}$. It is clear that $E_r(\rho)=0$
for the states with $p\leq\frac{2j}{2j+1}$. For this family of RI
states the set of separable states is just interval and the
definition of REE requires a minimization over this interval. Thus
for RI states of spin-$\frac{1}{2}$ and spin-$j$ particles the
minimizing separable state is the boundary state with
$p=\frac{2j}{2j+1}$. It follows from equation (\ref{red}) that
\begin{equation}
E_r(\rho)=\left\{
\begin{array}{ll}
0, & p\leq\frac{2j}{2j+1},\\
p\ln(\frac{2j+1}{2j}p)+(1-p)\ln[(2j+1)(1-p)], & p>\frac{2j}{2j+1},
\end{array}
\right.
\end{equation}
for a bipartite system of a spin-$\frac{1}{2}$ particle and a
spin-$j$ particle. We plot the $E_r(\rho)$ for $j=\frac{1}{2},\ 1,\
\frac{3}{2}$ in Figure \ref{2n} which shows us that the RI states
become less entangled as $j$ increase. K.K. Manne and C.M. Caves
obtained the same conclusion according to EoF in \cite{manne}.

\begin{figure}[h]
\begin{center}
\includegraphics{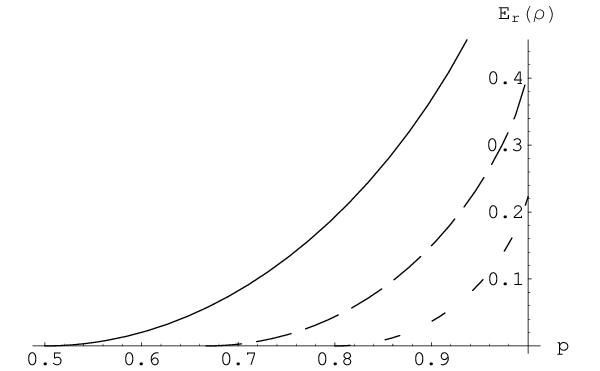}
\caption{The entanglement of relative entropy for $j=\frac{1}{2}$
(solid), $j=1$ (long-dashed) and $j=\frac{3}{2}$
(short-dashed).}{\label{2n}}
\end{center}
\end{figure}

Interestingly, we find that our result for two spin $\frac{1}{2}$
particles coincides with the case of two spin $\frac{1}{2}$
particles (i.e. Werner states) with the asymptotic value obtained by
K. Audenaert {\it et al.} in \cite{aude1}.

\section{REE for $3\otimes N$ system}\label{3xn}
In the section we discuss the REE of RI states of $j_1=1(N_1=3)$
particle and arbitrary $j_2$ particle, that is, of $3\otimes N_2$
system. Set $j=j_2$ and $N=N_2=2j_2+1$ for convenience. Since $J$
takes on the values $J=j-1,\ j\hbox{ and }j+1,\ \vec{\alpha}$ is a
three-vector $\vec{\alpha}=(\alpha_{j-1}\ \ \alpha_{j}\ \
\alpha_{j+1})^T.$ From equation (\ref{con}) we know the set of RI
states is given by the relations: $\alpha_{j-1},\ \alpha_{j},\
\alpha_{j+1}\geq0$ and
\begin{equation}\label{3n}
 \sqrt{\frac{N-2}{3N}}\alpha_{j-1}+\sqrt{\frac{1}{3}}\alpha_{j}
+\sqrt{\frac{N+2}{3N}}\alpha_{j+1}=1.
\end{equation}
We infer from the equation (\ref{con})  that $S^\alpha$ is a
2-simplex, i.e. a triangle with vertices as follows:
\begin{equation}\label{ABC}
  A=\left(
\begin{array}{c}
0\\
0\\
\sqrt{\frac{3N}{N+2}}
\end{array} \right),\ \ \ \
B=\left(
\begin{array}{c}
\sqrt{\frac{3N}{N-2}}\\
0\\
0
\end{array} \right),\ \ \ \
C=\left(
\begin{array}{c}
0\\
\sqrt{3}\\
0
\end{array} \right).
\end{equation}
In \cite{breuer1,breuer2} H.P. Breuer introduced the time reversal
transformation $\vartheta$ which is unitarily equivalent to the
transposition $T$. Therefore the Peres-Horodecki criterion can be
expressed by $\vartheta_2\rho=(I\otimes\vartheta)\rho\geq0.$ It is
worth mentioning that $\vartheta_2$ is taken to be of the form
$\vartheta_2(A\otimes B)=A\otimes\vartheta B=A\otimes
VB^TV^\dagger,$ where $V$ is a unitary matrix which represents a
rotation of the coordinate system about the $y$-axis by the angle
$\pi$. It follows from \cite{breuer2} that $\vartheta_2 S^\alpha$ is
also a 2-simplex with vertices
$$
A'=\left(
\begin{array}{c}
\sqrt{\frac{3(N-2)}{N}}\\
\frac{2\sqrt{3}}{N+1}\\
\frac{2}{N+1}\sqrt{\frac{3}{N(N+2)}}
\end{array} \right),\ \ \ \
B'=\left(
\begin{array}{c}
\frac{2}{N-1}\sqrt{\frac{3}{N(N-2)}}\\
-\frac{2\sqrt{3}}{N-1}\\
\sqrt{\frac{3(N+2)}{N}}
\end{array} \right),\ \ \ \
C'=\left(
\begin{array}{c}
-\frac{2}{N-1}\sqrt{\frac{3(N-2)}{N}}\\
\sqrt{3}\ \frac{N^2-5}{N^2-1}\\
\frac{2}{N+1}\sqrt{\frac{3(N+2)}{N}}
\end{array} \right).
$$
which are the images of $A,\ B$ and $C$ under the action of
$\vartheta_2$, respectively. Relation (\ref{3n}) implies that we can
represent the RI states using points in $\alpha$-space by two
coordinates $(\alpha_{j-1},\ \alpha_j)$. Consequently,
$S^\alpha_{PPT}(=S^\alpha\cap\vartheta_2 S^\alpha)$ is a polygon
with four vertices $A,\ A',\ D$ and $E$, where $A,\ A'$ are given by
the above equations and
\begin{equation*}
  D=\left(
\begin{array}{c}
\frac{N-1}{2}\sqrt{\frac{3}{N(N-2)}}\\
0\\
\frac{N+1}{2}\sqrt{\frac{3}{N(N+2)}}
\end{array} \right),\qquad
E=\left(
\begin{array}{c}
0\\
\sqrt{3}\ \frac{N-1}{N+1}\\
\frac{2}{N+1}\sqrt{\frac{3N}{N+2}}
\end{array} \right).\ \ \ \
\end{equation*}
H.P. Breuer \cite{breuer2} proved that PPT is necessary and
sufficient for separability of all $3\otimes N$ systems with odd
$N$. So the polygon $ADA'E$ represents the set of separable states
for a bipartite system of spin-1 particle and spin-$j$ particle with
integer $j$. Thus the separable state that minimizes any RI state of
spin-1 particle and spin-$j$ particle with integer $j$ can be
obtained. In the following we will discuss the REE for RI states of
$3\otimes N$ systems. Throughout this section we refer to the state
corresponding to the point $A$ as $\rho_A$ ($A$ stands for any
letter).

\subsection{The case $3\otimes3$}
\begin{figure}[h]
\begin{center}
\includegraphics{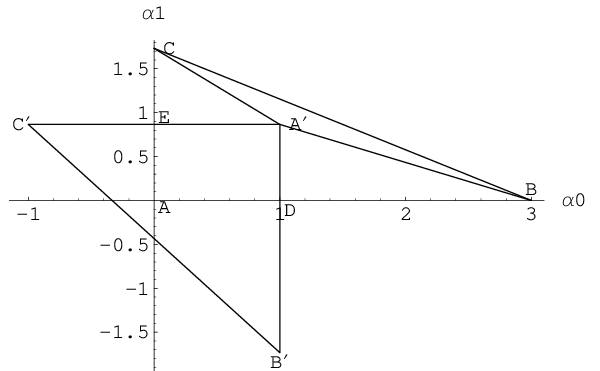}
\caption{$3\otimes3$ rotationally invariant states}{\label{33}}
\end{center}
\end{figure}

The state space for $3\otimes3$ RI states is split naturally into
four regions: the separable rectangle $ADA'E$ and the three
triangles $A'CE,\ A'BD$ and $A'BC$ in Figure \ref{33}.

It is obvious that $E_r(\rho)=0$ for any state in the rectangle
$ADA'E$. It is easy to see that the state $\rho_C$ and the states on
the whole line $EA'$ are depicted by the vector $\vec{\alpha}=(0,\
\sqrt{3},\ 0)^T$ and $\vec{\alpha}=(x,\ \frac{\sqrt{3}}{2},\
\frac{3}{\sqrt{5}}(\frac{1}{2}-\frac{x}{3}))^T$, respectively.
According to equation (\ref{red}), one can easily show any state on
the whole line connecting the points $A'$ and $E$, is a separable
state which minimizes the REE for $\rho_C$. As a result of the
property (5) of REE, we can find the minimizing separable state for
any state in the whole triangle $A'CE$. One just has to connect the
point of given states with the point $C$ to draw a straight line.
The intersection with the line $EA'$ is a minimizer for $\rho_C$ and
all states on the connecting line. Similarly, we can obtain the
minimizing separable states for $\rho_B$ and any state in the whole
triangle $A'BD$. In addition, all the states in the whole triangle
$A'BC$ have the same minimizer $\rho_{A'}$. In fact, one just has to
show that the minimizer for any state on the line $BC$ is the
separable state $\rho_{A'}$ because of the property (5) of relative
entropy of entanglement. In order to simplify the calculation, we
write any $3\otimes3$ rotationally invariant state which is depicted
by the vector $\vec{\alpha}=(3\alpha_0,\ \sqrt{3}\alpha_1,\
\frac{3}{\sqrt{5}}(1-\alpha_0-\alpha_1))^T$ as $\rho$, where
$\alpha_0\in[0,\ 1],\ \alpha_1\in[0,\ 1].$ Consequently, for RI
state $\rho$ in the system of two spin-1 particles we have

\begin{itemize}
\item If $\rho$ is in the rectangle $ADA'E$, then $E_r(\rho)=0$.
\item If $\rho$ is in the triangle $A'CE$, then $E_r(\rho)=
(1-\alpha_1)\ln[2(1-\alpha_1)]+\alpha_1\ln(2\alpha_1)$.
\item If $\rho$ is in the triangle $A'BC$, then
$$E_r(\rho)=
\alpha_0\ln(3\alpha_0)+\alpha_1\ln(2\alpha_1)+
(1-\alpha_0-\alpha_1)\ln[6(1-\alpha_0-\alpha_1)].$$
\item If $\rho$ is in the triangle $A'BD$, then $E_r(\rho)
=\alpha_0\ln(3\alpha_0)+(1-\alpha_0)\ln[\frac{3}{2}(1-\alpha_0)].$
\end{itemize}

\subsection{The case $3\otimes N$ with odd $N$}\label{case3n}
\begin{figure}[h]
\begin{center}
\includegraphics{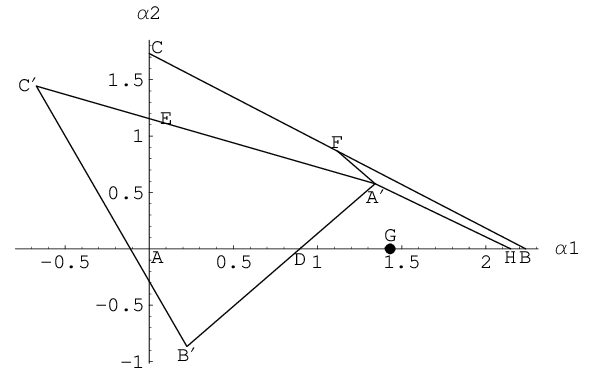}
\caption{$3\otimes5$ rotationally invariant states}{\label{35}}
\end{center}
\end{figure}

The state space for $3\otimes5$ RI states is split into four regions
by the below discussions for REE: the separable polygon $ADA'E$, the
entangled triangle $A'DH$, the entangled polygon $A'HBF$ and the
entangled polygon $A'FCE$ in figure \ref{35}. The coordinates of the
point $F,\ G$ and $H$ are $(\frac{\sqrt{5}}{2},\frac{\sqrt{3}}{2}),\
(\frac{16\sqrt{5}}{25},0)$ and $(\frac{24\sqrt{5}}{25},0)$,
respectively. Obviously, $E_r(\rho)=0$ for any state in the polygon
$ADA'E$.

One can write the states on the lines $BC,\ A'E$ and $A'D$ using the
parameter vectors
\begin{equation*}
 \vec{\alpha}=\left(
\begin{array}{c}
a\\
-\sqrt{\frac{3}{5}}\cdot a+\sqrt{3}\\
0
\end{array} \right),\
\vec{\alpha}_x=\left(
\begin{array}{c}
x\\
-\frac{\sqrt{15}}{9}x+\frac{2}{\sqrt{3}}\\
\sqrt{\frac{15}{7}}(\frac{1}{3}-\frac{4\sqrt{5}}{45}x)
\end{array} \right),\
\vec{\alpha}_y=\left(
\begin{array}{c}
y\\
\frac{\sqrt{15}}{3}y-\frac{2}{\sqrt{3}}\\
\sqrt{\frac{15}{7}}(\frac{5}{3}-\frac{8\sqrt{5}}{15}y)
\end{array} \right),
\end{equation*}
respectively. We denote the corresponding state to any point on the
whole line $BC$ as $\rho_a$ with $a\in[0,\sqrt{5}]$. Then from
equation (\ref{red}) we can obtain $E_r(\rho_a)=\min\{f(x),\
g(y)\},$ where
$$\begin{array}{rl}
f(x)&=\frac{a}{\sqrt{5}}(\ln a-\ln
x)+(-\frac{a}{\sqrt{5}}+1)[\ln(-\sqrt{\frac{3}{5}}\cdot
a+\sqrt{3})-\ln(-\frac{\sqrt{15}}{9}x+\frac{2}{\sqrt{3}})],\\
g(y)&=\frac{a}{\sqrt{5}}(\ln a-\ln
y)+(-\frac{a}{\sqrt{5}}+1)[\ln(-\sqrt{\frac{3}{5}}\cdot
a+\sqrt{3})-\ln(\frac{\sqrt{15}}{3}y-\frac{2}{\sqrt{3}})],
\end{array}$$
with $x\in[0,\frac{3}{\sqrt{5}}],\
y\in[\frac{2}{\sqrt{5}},\frac{3}{\sqrt{5}}]$. Here $f(x)$ is a
continuous function of $x$. Hence there must exist minimum for
function $f(x)$ in the closed interval $[0,\frac{3}{\sqrt{5}}]$.
Analogous to $f(x)$, there exists minimum for function $g(y)$ in the
closed interval $[\frac{2}{\sqrt{5}},\frac{3}{\sqrt{5}}]$. By a
tedious calculation, we can obtain
\begin{equation*}
E_r(\rho_a)=\left\{
\begin{array}{ll}
f(\frac{6a}{5}), & a\in[0,\frac{\sqrt{5}}{2}],\\
f(\frac{3}{\sqrt{5}}), & a\in[\frac{\sqrt{5}}{2},\sqrt{5}].
\end{array}
\right.
\end{equation*}

Every state on the line $BD$ can be represented by the parameter
vector $\vec{\alpha}=(b,\ 0,\
\sqrt{\frac{15}{7}}(1-\frac{b}{\sqrt{5}}))^T$. In the sequel, we
label the corresponding state to each point on the whole line $BC$
as $\rho_b$ with $b\in[\frac{2}{\sqrt{5}},\sqrt{5}]$. Then
$E_r(\rho_b)=\min\{f(x),\ g(y)\},$ here
$$\begin{array}{rl}
f(x)&=\frac{b}{\sqrt{5}}(\ln b-\ln
x)+(-\frac{b}{\sqrt{5}}+1)\{\ln[\sqrt{\frac{15}{7}}(-\frac{b}{\sqrt{5}}+1)]
-\ln[\sqrt{\frac{15}{7}}(\frac{1}{3}-\frac{4\sqrt{5}}{45}x)]\},\\
g(y)&=\frac{b}{\sqrt{5}}(\ln b-\ln
y)+(-\frac{b}{\sqrt{5}}+1)\{\ln[\sqrt{\frac{15}{7}}(-\frac{b}{\sqrt{5}}+1)]
-\ln[\sqrt{\frac{15}{7}}(\frac{5}{3}-\frac{8\sqrt{5}}{15}y)]\},
\end{array}$$
with $x\in[0,\frac{3}{\sqrt{5}}],\
y\in[\frac{2}{\sqrt{5}},\frac{3}{\sqrt{5}}]$. By a similar
derivation, we can obtain
\begin{equation*}
E_r(\rho_b)=\left\{
\begin{array}{ll}
g(\frac{5b}{8}), & b\in[\frac{16\sqrt{5}}{25},\frac{24\sqrt{5}}{25}],\\
g(\frac{3}{\sqrt{5}}), & b\in[\frac{24\sqrt{5}}{25},\sqrt{5}].
\end{array}
\right.
\end{equation*}

To summarize, we present that the separable states corresponding to
the point
$P=(\frac{6a}{5},-\frac{2a}{\sqrt{15}}+\frac{2}{\sqrt{3}})$ and
$Q=(\frac{5b}{8},\frac{5\sqrt{15}}{24}b-\frac{2}{\sqrt{3}})$
minimize the states $\rho_a$ with $a\in[0,\frac{\sqrt{5}}{2}]$ and
$\rho_b$ with $b\in[\frac{16\sqrt{5}}{25},\frac{24\sqrt{5}}{25}]$,
respectively. Accordingly, it is a remarkable fact that the state
$\rho_{A'}$ is the minimizing separable state for the states
$\rho_a$ with $a\in[\frac{\sqrt{5}}{2},\sqrt{5}]$ and $\rho_b$ with
$b\in[\frac{24\sqrt{5}}{25},\sqrt{5}]$. We should emphasize that the
states on the whole line $DG$ have the same minimizing state
$\rho_D$. The corresponding nearest separable state on the line
$A'D$ approaches $\rho_D$ as the state on the line $HG$ approaches
$\rho_G$. Hence, one may
 take the triangle $A'DH$ as the polygon $A'DGH$.
According to the property (5) of REE, we can obtain the REE for any
$3\otimes5$ RI state (see the below paragraph).

In much the same way as the above derivation, in terms of relative
entropy of entanglement, the state space for $3\otimes N$ with odd
$N$ RI states is split into four regions: the separable polygon
$ADA'E$, the entangled triangle $A'DH$, the entangled polygon
$A'HBF$ and the entangled polygon $A'FCE$. The coordinates of $F,\
G$ and $H$ are $(\frac{N-3}{N-1}\sqrt{\frac{3N}{N-2}},\
\frac{2}{N-1}\sqrt{3}),\
(\frac{(N-1)^2(N+3)}{2(N^2-5)}\sqrt{\frac{3}{N(N-2)}},\ 0)$ and
$(\frac{(N+3)(N-1)}{N^2-5}\sqrt{\frac{3(N-2)}{N}},\ 0),$
respectively. It is easy to see that $E_r(\rho)=0$ for any state in
the polygon $ADA'E$. In addition, $\rho_{A'}$ is the minimizing
state for any state of the polygon $A'HBF$. By a similar discussion
with the case $3\otimes5$, we can obtain that the coordinates of $P$
and $Q$ are $(\frac{(N-2)(N-1)}{N(N-3)}a,\
\frac{N-1}{N+1}(-\sqrt{\frac{N-2}{N}}a+\sqrt{3}))$ and
$(\frac{N^2-5}{(N+3)(N-1)}b,\
\frac{4(N^2-5)\sqrt{N(N-2)}}{(N^2-1)(N^2-9)}b
-\frac{2\sqrt{3}(N-1)}{(N+1)(N-3)})$, respectively. All the states
on the line connecting the point $(a,-\sqrt{\frac{N-2}{N}}\cdot
a+\sqrt{3})$ with the point $P$ have the same minimizing separable
state $\rho_P$ in the polygon $A'CEF$. Similarly, all the states on
the line connecting the point $(b,0)$ with the point $Q$ have the
same minimizing separable state $\rho_Q$ in the triangle $A'DH$. It
is worth mentioning that the separable state $\rho_D$ minimizes all
the states on the whole line $DG$. To simplify the calculation,
suppose that an RI state of spin-1 particle and spin-$j$ particle
with integer $j$ is depicted by the vector
$\vec{\alpha}=(\sqrt{\frac{3N}{N-2}}\alpha_{j-1},\
\sqrt{3}\alpha_j,\
\sqrt{\frac{3N}{N+2}}(1-\alpha_{j-1}-\alpha_j))^T$, where
$\alpha_{j-1}\in[0,\ 1],\ \alpha_j\in[0,\ 1].$ Consequently, for RI
state $\rho$ in the system of a spin-1 particle and a spin-$j$
particle with integer $j>1$, we have

\begin{itemize}
\item If $\rho$ is in the rectangle $ADA'E$, then $E_r(\rho)=0$.
\item If $\rho$ is in the triangle $A'FCE$, then
$$\begin{array}{rl}
E_r(\rho)=&
\alpha_{j-1}\ln\frac{N(N-3)\alpha_{j-1}}{(N-1)(N-2)a}+\alpha_j\ln
\frac{(N+1)\alpha_j}{(N-1)(1-a)}\\[4mm]
&+(1-\alpha_{j-1}-\alpha_j)\ln\frac{N(N+1)(N-3)(1-\alpha_{j-1}-\alpha_j)}
{2[N(N-3)-(N-1)^2a]},
\end{array}$$
here
$$a=\frac{-t_1-\sqrt{t_1^2-4N(N-1)^2(N-3)\alpha_{j-1}}}{2(N-1)^2},$$
$$t_1=(N+1)\alpha_j+N(N-3)\alpha_{j-1}-(N-1)^2.$$
\item If $\rho$ is in the triangle $A'HBF$,\\
$E_r(\rho)=
\alpha_{j-1}\ln(\frac{N}{N-2}\alpha_{j-1})+\alpha_j\ln(\frac{N+1}{2}\alpha_j)
+(1-\alpha_{j-1}-\alpha_j)\ln[\frac{N(N+1)}{2}(1-\alpha_{j-1}-\alpha_j)].$
\item If $\rho$ is in the triangle $A'DH$,\\[3mm]
$$\begin{array}{rl}
E_r(\rho)
=&\alpha_{j-1}\ln\frac{(N+3)(N-1)\alpha_{j-1}}{(N^2-5)b}+\alpha_j\ln
\frac{(N^2-1)(N^2-9)\alpha_j}{4N(N^2-5)b-2(N+3)(N-1)^2}\\[4mm]
&+(1-\alpha_{j-1}-\alpha_j)\ln\frac{(N+1)(N-3)(1-\alpha_{j-1}-\alpha_j)}
{(N^2-5)(1-b)},
\end{array}$$
here
$$\begin{array}{rl}
b&=\frac{t_2+\sqrt{t^2-8N(N^2-5)(N-1)^2(N+3)\alpha_{j-1}}}{4N(N^2-5)},\\
t_2&=(N+3)(N-1)^2+2N(N^2-5)\alpha_{j-1}+(N+1)^2(N-3)\alpha_j.
\end{array}$$
\end{itemize}

\subsection{The case $3\otimes N$ with even $N$}
Since the set of the separable states is bounded by the straight
lines $AE,\ A'E$ and a concave curve for a bipartite system of
spin-1 particle and spin-$j$ particle with half-integer $j$
\cite{breuer2}, it is cumbersome to compute the relative entropy of
entanglement for this kind of RI states. However, we have some
interesting results on RI states of $3\otimes N$ system with even
$N$. Considering $E_\Gamma(\rho)$, the state space for $3\otimes N$
with even $N$ RI states is also split into four regions: the PPT
polygon $ADA'E$, the entangled triangle $A'DH$, the entangled
polygon $A'HBF$ and the entangled polygon $A'FCE$.

Analogous to theorem 4 in \cite{vedral2}, one can reduce to the
property of $E_\Gamma(\rho)$: if $\sigma^*$ minimizes
$S(\rho\|\sigma^*)$ over $\sigma\in{\mathcal P}$ then $\sigma^*$ is
also a minimum for any state of the form
$\rho_x=(1-x)\rho+x\sigma^*.$ Thus making use of this property of
$E_\Gamma(\rho)$ and the similar derivation to relative entropy of
entanglement for RI state of $3\otimes5$ system, we obtain
$E_\Gamma(\rho)$ for all rotationally invariant states of a spin-1
particle and a spin-$j$ particle with half-integer $j$.
$E_\Gamma(\rho^\prime)$ for RI state $\rho^\prime$ of $3\otimes N$
system with even $N$ has the same expression as $E_r(\rho)$ for RI
state $\rho$ of $3\otimes N$ system with odd $N$ in subsection
\ref{case3n}. According to the property (8) of relative entropy,
$E_\Gamma(\rho)$ provides us a lower bound of relative entropy of
entanglement and an upper bound on the rate at which entanglement
can be distilled.

\section{Conclusions}\label{conclu}
It was argued that the REE is the most appropriate quantity to
measure distinguishability between different quantum states. Hence
it could be a powerful tool for investigating quantum channels¡¯
properties\cite{vedral3}. In the present paper, we give the formula
of the REE for RI states of $2\otimes M$ system and $3\otimes N$
system with odd $N$. RI states of spin-1 particle and spin-$j$
particle constitute a two-parameter family. Therefore, relative
entropy of entanglement for them is a function of two variables. One
can find that the expression of $E_r(\rho)$ for RI states of
$3\otimes N$ system with odd $N$ can not be applied to the RI states
of two spin-1 particles. It shows that RI states of two equal spins
characterize distinct entanglement from RI states of two different
spins. Although EoF for RI states is difficult to compute, we give a
lower bound of EoF for RI states of spin-1 and arbitrary spin-$j$
particles. Meanwhile, it is an upper bound for the number of singlet
states that can be distilled from a given RI state. In addition,
H.P. Breuer \cite{breuer2} points out that $S_{PPT}^\alpha$
approaches the set $S^\alpha$ meanwhile
$S^\alpha_{sep}(=S^\alpha\cap\mathcal D)$ approaches
$S_{PPT}^\alpha$ as $N$ increases. Thus we find that the relative
entropy of entanglement vanishes for $3\otimes N$ RI states when
$N\rightarrow \infty.$ Interestingly, the asymptotic relative
entropy of entanglement with respect to positive partial transpose
(AREEP) which is defined as the regularisation
\begin{eqnarray*}
E^\infty_r(\rho)=\lim_{n\rightarrow\infty}\frac{1}{n}E_r(\rho^{\otimes
n}),
\end{eqnarray*}
is investigated on Werner states and orthogonally invariant state as
a sharper bound to distillable entanglement in \cite{aude1,aude2}.
Therefore, we will investigate the AREEP on RI states in the future
work.

\section{Acknowledgement} The authors would like to appreciate the
referees' valuable suggestions. This work is supported by the NNSF
of China (Grant No. 10871227).


\begin{thebibliography}{20}

\section*{References}

\bibitem{nilsen} M.A. Nilsen and I.L. Chuang, {\it Quantum Computation
and Quantum Information}, Cambridge University Press, Cambridge,
2002.

\bibitem{hill} S. Hill and W.K. Wootters, Phys. Rev. Lett.
78 (1997) 5022.

\bibitem{woot} W.K. Wootters, Phys. Rev. Lett.  80
(1998) 2245.

\bibitem{rain1} E.M. Rains, Phys. Rev. A 60 (1999) 179.

\bibitem{chen} Y.X. Chen and D. Yang, Quan. Inf. Proc. 1
(2003) 389.

\bibitem{caves} P. Rungta and C.M. Caves, Phys. Rev. A 67
(2003) 012307.

\bibitem{ck} K. Chen, S. Albeverio and S.M. Fei, Reports on Math.
Phys.  58 (2006) 449.

\bibitem{john1} J. Schliemann, Phys. Rev. A 68 (2003) 012309.

\bibitem{john2} J. Schliemann, Phys. Rev. A 72 (2005) 012307.

\bibitem{breuer1} H. P. Breuer, Phys. Rev. A 71 (2005) 062330.

\bibitem{breuer2} H.P. Breuer, J. Phys. A: Math. Gen. 38
(2005) 9019.

\bibitem{breuer3} H.P. Breuer, Phys. Rev. Lett. 97
(2006) 08050.

\bibitem{chru3} D. Chruscinski and A. Kossakowski, Open Sys. Inf. Dyn. 14
(2007) 25.

\bibitem{augu} R. Augusiak and J. Stasi\'nska, Phys. Lett. A
363 (2007) 182.

\bibitem{manne} K.K. Manne and C.M. Caves, Quan. Inf. Comp.
8 (2008) 0295.

\bibitem{werner} K.G.H. Vollbrecht and R.F. Werner, Phys. Rev. A
64 (2002) 062307.

\bibitem{vedral1} V. Vedral, M.B. Plenio, M. A. Rippin and P. L.
Knight, Phys. Rev. Lett. 78 (1997) 2275.

\bibitem{vedral2} V. Vedral and M.B. Plenio, Phys. Rev. A 57
(1996) 1619.

\bibitem{werner2} R.F. Werner, Phys. Rev. A 40 (1989) 4277.

\bibitem{vedral3} V. Vedral, Rev. Mod. Phys. 74 (2002) 197.

\bibitem{aude1} K. Audenaert, J. Eisert, E. Jan\'e, M.B. Plenio, S.
Virmani and B. De Moor, Phys. Rev. Lett 87 (2001) 217902.

\bibitem{aude2} K. Audenaert, B. De Moor, K.G.H. Villbrecht and R.F.
Werner, Phys. Rev. A 66 (2002) 032310.

\end{thebibliography}
\end{document}